\documentclass[11pt]{article}
\usepackage{amsmath,amssymb,amsfonts}
\usepackage{latexsym}
\usepackage{graphicx}
\usepackage{epsfig}

\begin{document}


\begin{centering}
 
{\ }\vspace{0.5cm}
 
{\Large\bf Extended two-level quantum dissipative system from}

\vspace{5pt}

{\Large\bf   bosonization of the elliptic spin-$\frac12$ Kondo model}

\vspace{0.5cm}

Sol H. Jacobsen\footnote{Commonwealth Endeavour Scholar} and
P. D. Jarvis\footnote{Alexander von Humboldt Fellow}${}^,$\footnote{Tasmanian Senior Fulbright Scholar}

\vspace{0.3cm}

{\em School of Mathematics and Physics, University of Tasmania}\\
{\em Private Bag 37, Hobart, Tasmania 7001, Australia }\\
{\em E-mail: {\tt solj@utas.edu.au},  {\tt Peter.Jarvis@utas.edu.au}}\\
{\em{Journal Ref:} J. Phys. A: Math. Theor. 44 (2011) 115003.}

\end{centering} 
\section*{Abstract}
We study the elliptic spin-$\frac 12$ Kondo model (spin-$\frac 12$ fermions in one dimension with fully anisotropic contact interactions with a magnetic impurity) in the light of mappings to bosonic systems using the fermion-boson correspondence and associated unitary transformations. We show that for fixed fermion number, the bosonic system describes a two-level quantum dissipative system with two noninteracting copies of infinitely-degenerate upper and lower levels. In addition to the standard tunnelling transitions, and the transitions driven by the dissipative coupling, there are also bath-mediated transitions between the upper and lower states which simultaneously effect shifts in the horizontal degeneracy label. We speculate that these systems could provide new examples of continuous time quantum random walks, which are exactly solvable. 


\noindent





\setcounter{footnote}{0}
\section{Introduction}
\label{sec:Introduction}
The importance of the Kondo models in condensed matter physics has been well publicised since the
original Kondo model was first used to describe electron resistivity in dilute magnetic alloys \cite{Kondo1964}. The field
of applicability was considerably extended when it was shown that resolving the dilute Kondo problem was equivalent
to studying the physics of the ground state of a spin-$\frac 12$ magnetic impurity interacting with a free electron gas \cite{AndersonandYuval1969}. Although these examples firmly established the model's numerical predictive power, it was not until a decade later that the model's true significance in providing exact results became apparent. It is now well known that the one-dimensional spin-$\frac 12$ Kondo Hamiltonians are exactly solvable \cite{Baxter1982,TsvelickandWiegmann1983}, through the mapping of the fermion-impurity scattering matrices to the
$R$-matrices that satisfy the Yang-Baxter equation, and the associated Bethe \emph{Ansatz} for their eigenstates \cite{Bethe1931}. The importance and applicability
of the Kondo models was again extended after it became established that the standard,
spin-$\frac 12$ \emph{anisotropic} Kondo model can be mapped to a particular case of the spin-$\frac 12$ spin-boson model -- the simplest possible quantum dissipative system (QDS) of a two-level quantum system with 
so-called ohmic coupling to an oscillator bath \cite{Leggettetal1987,Weiss2008}. We have shown recently in \cite{JacobsenJarvis2010} that re-examination of suitable \emph{multicomponent} fermion gas-impurity models mapped to bosonic systems via the method of constructive bosonisation \cite{Haldane1981, vonDelftandSchoeller1998} and unitary transformation leads to new instances of exactly solvable quantum dissipative systems beyond the two-level QDS. 

In this brief report we will demonstrate that the same method can be used to reveal a new intricate structure of a previously well-studied system in the hierarchy of exactly-solvable models. Namely, beyond the standard anisotropic Kondo model with its magnetic spin-spin couplings of `$X\!X\!Z$'-type and residual axial symmetry, lies the \emph{fully} anisotropic Kondo model \cite{Baxter1982,TsvelickandWiegmann1983,TakhtadzhanandFaddeev1979} with $XY\!Z$-type interaction. As is well known, in the related language of quantum spin chains, the $X\!X\!Z$ model entails trigonometric $R$-matrices for its solution, whereas the $XY\!Z$ model requires $R$-matrices parametrised with elliptic functions, and the associated Bethe \emph{Ansatz} equations for them.

We shall briefly review the $XY\!Z$-type Kondo model and the methods of constructive bosonisation and
unitary transformation in \S 2, before discussing the details of the resulting structure in \S 3. In particular,
we show in detail that for $N$ fermions, the associated QDS corresponds to an oscillator bath in interaction with an \emph{extended} two-level system, where the upper and lower levels have infinite degeneracy. There is an additional quantum number, inherited from the fermion-boson correspondence, which can be used to label states in a `horizontal' direction (see figure \ref{Lattice}).  In \S 4 by way of conclusions to this brief report, we further discuss this structure to highlight the roles of specific model components and the emergence of bath-mediated tunnelling transitions, which are not present in the original spin-boson model, and importantly also to suggest potential new applications -- in particular in relation to models of quantum random walks.

\section{Re-examination of $XYZ$-type Kondo model, bosonisation and unitary transformation}
\label{sec:XYZ}
The fully anisotropic Kondo model derives from of the original isotropic Kondo Hamiltonian (in three dimensions),
\begin{eqnarray}
H_K=\sum_{\mathbf{k},\alpha}\epsilon(\mathbf{k})c^\dagger_{\mathbf{k}\alpha}c_{\mathbf{k}\alpha} + J\boldsymbol{S}\cdot\boldsymbol{s}(0).\label{Kondoeq}
\end{eqnarray}
The mode operators $c{}^\dagger_{{\mathbf k}\alpha}$, $c_{{\mathbf k}\alpha}$ respectively create and annihilate fermion states with spin label\footnote{The labels $\alpha=1,2$ or $\uparrow$, $\downarrow$ denote the eigenvalues $+1$, $-1$ of $\sigma_z$, respectively.}  $\alpha$ and wave vector ${\mathbf k}$. The impurity spin generators are ${\mathbf S} =\textstyle{\frac 12} \mbox{\boldmath{$\sigma$}}$ (\mbox{\boldmath{$\sigma$}}  being the vector of $2\!\times\!2$ Pauli matrices), and the notation
${\mathbf s}(\mathbf{0}) := \sum_{{\mathbf k},{\mathbf k}', \alpha, \alpha'} c{}^\dagger_{{\mathbf k}\alpha}
(\textstyle{\frac 12}\mbox{\boldmath{$\sigma$}})_{\alpha \alpha'} c_{{\mathbf k}'\alpha'} $ indicates that the impurity couples to the fermion spin density at the origin. The parameter $J$ has conventional dimensions of energy, with positive sign, $J>0$, in the antiferromagnetic case. Generalisations of the coupling term dictate the anisotropy of this class of models. A natural nomenclature is to refer to such models as being of $X\!X\!X$-type for the isotropic case, of $X\!X\!Z$-type for the standard anisotropic case, and of $XY\!Z$-type for the fully anisotropic case that we shall discuss here. 

For weak coupling, small interaction amplitudes mean that dominant excitations are close to the Fermi surface for long times and low temperatures, such that the dispersion relation can be linearised about the Fermi energy,
$\varepsilon({\mathbf k}) = \varepsilon_F + \hbar v_F(|{\mathbf k}|-k_F)$   (where $v_F$ is the Fermi velocity and $k_F$ is the Fermi wavenumber. Furthermore, under the assumption that $s$-wave scattering is the dominant process in (1), the model becomes essentially one-dimensional. The fully anisotropic $XY\!Z$-type Kondo model that we consider is therefore
\begin{eqnarray}
H^{XY\!Z}&=& \hbar v_F\sum_{\widetilde{k},\alpha=1,2}\widetilde{k}c^\dagger_{\widetilde{k}\alpha}c_{\widetilde{k}\alpha} + \frac{J_x}{2}\sum_{\widetilde{k},\widetilde{k'}}\left(c^\dagger_{\widetilde{k}\uparrow}c_{\widetilde{k'}\downarrow}+c^\dagger_{\widetilde{k}\downarrow}c_{\widetilde{k'}\uparrow}\right) S_x\nonumber \\
&&+ \frac{J_y}{2i}\sum_{\widetilde{k},\widetilde{k'}}\left(c^\dagger_{\widetilde{k}\uparrow}c_{\widetilde{k'}\downarrow} - c^\dagger_{\widetilde{k}\downarrow}c_{\widetilde{k'}\uparrow}\right) S_y +J_z\sum_{\widetilde{k},\widetilde{k'}}\left(c^\dagger_{\widetilde{k}\uparrow}c_{\widetilde{k'}\uparrow} - c^\dagger_{\widetilde{k}\downarrow}c_{\widetilde{k'}\downarrow}\right) S_z.\label{XYZKondo}
\end{eqnarray}
Here $|{\mathbf k}|-k_F$ has been replaced by a one-dimensional wavenumber $\widetilde{k}\ge -k_F$, and we have introduced three independent quantities $J_x$, $J_y$, $J_z$ to parametrise the (in general unequal) impurity-spin interactions in the different coordinate directions. 

As mentioned above, the mapping between the $X\!X\!Z$-type anisotropic Kondo model (the case $J_x=J_y$) and a two-level quantum dissipative system is well known \cite{Leggettetal1987,CostiandZarand1999}, and relies on the procedure of bosonisation and unitary transformation as we shall use below. Re-examination of the corresponding mapping for the $XY\!Z$-type, fully anistropic model reveals a connection to an extended two-level dissipative system, with some structural peculiarities that have not been highlighted in the literature which we wish to address here.

\subsection{Bosonisation}
\label{subsec:Bosonisation}
The fermion-boson correspondence \cite{Leggettetal1987, vonDelftandSchoeller1998} sets up an isomorphism between the fermionic state space and operators thereon, and associated bosonic Fock spaces. In the sequel it will be convenient to use Fourier expansions to form expressions for local quantum fields, both for spinors as well as the associated scalars. For example, 
\begin{eqnarray}
\label{eq:PsiPhiNoA}
\psi_\alpha(x) = & \, \sqrt{  {\frac{2\pi}{L}}} \sum_{k} e^{-ikx}c_{k\alpha},
\qquad \psi^\dagger_\alpha(x) =  \sqrt{  {\frac{2\pi}{L}}} \sum_k e^{ikx}c^\dagger_{k\alpha},
\end{eqnarray}
entails a sum over fermionic creation and annihilation modes for spin label $\alpha$ and wavenumber $k=2\pi n_k/L$, for sample size $L$, $0\le x < L$ and (for example) periodic boundary conditions with integral $n_k $. 
A fundamental observation is that the bilinear combinations
\begin{eqnarray}
\label{eq:bs}
b^\dagger_{p \alpha}  = & \, i\sqrt{  {\frac{2\pi}{Lp}}}\sum_{k=-\infty}^{\infty}\mbox{\boldmath{$:$}} c^\dagger_{k\!+\!p\, \alpha} c_{k\alpha}\mbox{\boldmath{$:$}}, \qquad 
b_{p \alpha}  = - i\sqrt{  {\frac{2\pi}{Lp}}}\sum_{k=-\infty}^{\infty} \mbox{\boldmath{$:$}} c^\dagger_{k\!-\!p \, \alpha} c_{k\alpha} \mbox{\boldmath{$:$}},
 \nonumber
\end{eqnarray}
where now $p= 2\pi n_p/L >0$, satisfy standard oscillator commutation relations, ${[}b_{p\alpha},b^\dagger_{p'\beta}{]}=\delta_{pp'}\delta_{\alpha\beta}$, ${[}b_{p\alpha},b_{p'\beta}{]}=0={[}b^\dagger_{p\alpha},b^\dagger_{p'\beta}{]}$ and so provide mode expansions for associated scalar fields, 
\begin{eqnarray}
\label{eq:PsiPhiA}
\varphi_{\alpha}(x) = & \, -\sum_{p>0} \sqrt{  {\frac{2\pi}{Lp}}} 
                  \big(e^{-ipx}  b_{p \alpha} \!+\! e^{ipx} b^\dagger_{p \alpha} \big).
\end{eqnarray}

Normal ordering is with respect to the fermionic vacuum defined by $c_{\alpha k}|0 \rangle = 0$, $k>0$; $c{}^\dagger_{\alpha k}| 0 \rangle = 0$, $k \le 0$. Clearly the $b_{\alpha p}$, $b{}^\dagger_{\alpha p}$ commute with total charge (fermion number) $N_\alpha = \sum_k \mbox{\boldmath{$:$}}c^\dagger_{k\alpha}c_{k\alpha}\mbox{\boldmath{$:$}}$.
The mapping of state spaces is such that to \emph{each} individual sub-sector of fixed charge, say $N_\alpha$, there is a copy of bosonic Fock space with a unique charge vacuum $| 0 \rangle_{N_\alpha}$ (a special case being $|0\rangle \equiv |0\rangle_0$). The fermion-boson correspondence at the level of the field operators is finally
\begin{eqnarray}
\label{eq:OperatorIdNoA}
\psi_\alpha(x) = & \, \sqrt{  {\frac{2\pi}{L}}}  {\mathcal F}_\alpha \,\mbox{\boldmath{$:$}} 
e^{-i\varphi_\alpha(x)}\mbox{\boldmath{$:$}},
\end{eqnarray}
where standard oscillator normal ordering is used. The so-called Klein operators ${\mathcal F}_\alpha$ are defined \cite{vonDelftandSchoeller1998} via sums over charge vacua of the form 
${\mathcal F}_\alpha\!\!=\!\!\sum_{N_\alpha}|0\rangle_{N_\alpha-1}\,\big(\!\!-1\big)^{\sum_{\alpha'=1}^{\alpha\!-\!1}N_{\alpha'}}\,{}_{N_\alpha}\!\langle 0|$
, with ${\mathcal F}{}^\dagger_\alpha$ correspondingly increasing the charge. The Klein factors are unitary operators satisfying 
\[
 {\mathcal F}_\alpha {\mathcal F}^\dagger_\alpha =  {\mathcal F}^\dagger_\alpha {\mathcal F}_\alpha =1, 
 \qquad 
 {\{} {\mathcal F}^\dagger_\alpha,{\mathcal F}_\beta{\}} = {\{} {\mathcal F}_\alpha, {\mathcal F}_\beta{\}} =
 {\{} {\mathcal F}^\dagger_\alpha,{\mathcal F}^\dagger_\beta{\}} = 0, \quad \alpha \ne \beta.
 \]
 
Turning to the comparison with (\ref{XYZKondo}), note that the momentum reference $\tilde{k}$ relative to the Fermi level is restricted to the range $\tilde{k}\ge -k_F$. However, it is clear that (\ref{eq:PsiPhiNoA}), (\ref{eq:PsiPhiA}) and (\ref{eq:OperatorIdNoA}) require a doubly infinitely-extended set of fermionic modes (with \emph{arbitrary} integer mode number $n_k$). In establishing the formal correspondence between the Kondo models and their
dissipative system counterparts the procedure is therefore to extend the limit on $\tilde{k}$ downwards to $-\infty$, equating $\tilde{k}$ in (\ref{XYZKondo}) with ${k}$ in (\ref{eq:PsiPhiNoA}), (\ref{eq:PsiPhiA}) and identifying the vacuum of the latter with the filled Fermi sea. As a consequence, the energy spectrum of the model as a whole is formally unbounded below, and requires the imposition of a cutoff. To maintain the rigorous mathematical fermion-boson correspondence and isomorphism
of Hilbert spaces throughout, this is achieved via the method of ``constructive bosonisation'' \cite{vonDelftandSchoeller1998}. This
introduces a regularisation parameter $a\rightarrow 0$ which sets a scale for the suppression of contributions from
wavenumbers $|p|\ge a^{-1}$ away from the Fermi surface, by modifying (\ref{eq:PsiPhiA}) above to 
\begin{eqnarray}
\label{eq:VarPhiWithA}
\varphi_{\alpha}(x) = & \, -\sum_{p>0} \sqrt{  {\frac{2\pi}{Lp}}} 
                  \big(e^{-ipx}  b_{p \alpha} \!+\! e^{ipx} b^\dagger_{p \alpha} \big){{ e^{-ap/2} }}.
\label{BosField1}
\end{eqnarray} An important technical consequence is that normal ordering in operator products can be re-expressed in terms of
ordinary products in an expansion in $a$. In particular, (\ref{eq:OperatorIdNoA}) becomes
\begin{eqnarray}
\psi_\alpha(x) = & \, \lim_{a\rightarrow 0}\left( \frac{{\mathcal F}_\alpha}{\sqrt{a}}  e^{-i\varphi_\alpha(x)}\right).\label{fermtoboson}
\end{eqnarray}

We can now express the various contributions to the fully anisotropic Kondo Hamiltonian $H^{XY\!Z}$ in terms of their bosonic counterparts (in the sense of a limit as in (\ref{fermtoboson})). Firstly, for the free fermion kinetic energy term in (\ref{XYZKondo}), a standard operator identity yields
\begin{eqnarray}
^B\!H_{kin} = \hbar v_F\int^{L/2}_{-L/2}\frac{dx}{2\pi}\frac12\boldsymbol{:}\left(\partial_x\varphi_\alpha(x)\right)^2\boldsymbol{:}.\label{Hkin}
\end{eqnarray}
It is useful to rearrange the interacting part of (\ref{XYZKondo}) into the form $H_{int} = H_\perp + H_\parallel+ H'$, where we introduce the new couplings $J_\parallel = J_z$, $J_\perp = \textstyle{\frac 12}(J_x+J_y)$, $J' = \textstyle{\frac 12}(J_x-J_y)$. Clearly the $X\!X\!Z$ case entails $J_x=J_y$ whence $J'=0$ -- the interaction becomes purely longitudinal and transverse, with no additional contribution of type $H'$. The bosonic counterparts using (\ref{eq:PsiPhiNoA}), (\ref{fermtoboson}) are
\begin{eqnarray}
\qquad \qquad ^B\!H_\perp + ^B\!\!H_\parallel \!\!\!&=& \frac{J_\bot L}{4\pi a}\left(e^{i\sqrt{2}\varphi_S(0)}S'_- + e^{-i\sqrt{2}\varphi_S(0)}S'_+\right)\nonumber\\
&& + \frac{J_{||} L}{2\pi}\sqrt{2}\partial_x\varphi_S(0) S_z , \nonumber \\ 
\mbox{whereas} \qquad \qquad ^B\!H' &=& \frac{J' L}{4\pi a}\left(e^{i\sqrt{2}\varphi_S(0)}S''_+ + e^{-i\sqrt{2}\varphi_S(0)}S''_-\right). \label{pretransf}
\end{eqnarray}
Here the spin combinations $\varphi_{S}(x) := (\varphi_1(x) - \varphi_2(x))/\sqrt{2}$, and $S_\pm'$, $S_\pm''$ are Klein operator-dressed spin raising and lowering operators, with $S_+' ={\mathcal F}_\downarrow{}^\dagger {\mathcal F}_\uparrow S_+$, 
$S_-' ={\mathcal F}_\uparrow{}^\dagger {\mathcal F}_\downarrow S_-$, $S_+'' ={\mathcal F}_\uparrow{}^\dagger {\mathcal F}_\downarrow S_+ $ and $S_-'' ={\mathcal F}_\downarrow{}^\dagger {\mathcal F}_\uparrow S_-$.

\subsection{Unitary transformation}
\label{subsec:UnitaryTransformation}
The final step in rearranging the model in bosonic form is to implement a simplifying unitary transformation. Specifically, the transformation $H \rightarrow U H U^{\dagger}$ with $U = \exp(-i\sqrt{2} \varphi_S(0) S_z)$ for the interaction part of the Hamiltonian in this case produces
\begin{eqnarray}
U \,^{\,B}\!H_{int}^{XYZ}\,U^{\dagger} &=& \frac{J'L}{4\pi a}\left( e^{i2\sqrt{2}\varphi_S(0)}S''_+ + e^{-i2\sqrt{2}\varphi_S(0)}S''_-\right)\nonumber\\
&&+ \frac{J_\bot L}{4\pi a}\left(S'_- + S'_+\right) + \frac{J_\parallel L\sqrt{2}}{2\pi}\partial_x\varphi_S(0) S_z.\label{XYZHamil}
\end{eqnarray}
Note that the similarity transformation removes the exponentials in the scalar fields in all but $H'$. For the case $J'=0$ we thus recover the interaction components of the standard spin-$\textstyle{\frac 12}$ spin-boson model,
\begin{eqnarray}
U \,^{\,B}H^{X\!X\!Z}_{int}\, U^{\dagger} = \frac{J_\bot L}{4\pi a}\left(S'_- + S'_+\right) + \frac{J_\parallel L\sqrt{2}}{2\pi}\partial_x\varphi_S(0) S_z.\nonumber
\end{eqnarray}
As detailed in \cite{JacobsenJarvis2010}, by gathering terms with the free Hamiltonian (\ref{Hkin}), the dissipative couplings $C_p :=  -\sqrt{2\alpha}\left(2\pi v_F\omega_p/L\right)^{1/2}e^{-\omega_p a/2v_F}$ make up the standard ohmic form of the spectral density for the interaction with the oscillator bath, with ohmic coupling constant $\alpha = \left(1- \left(J_\parallel L/2\pi\hbar v_F\right)\right)^2$. As the full Hamiltonian for the extended two-level system comprises the interaction plus the kinetic components $H=U\,^B\!H_{kin}U^\dagger +U\,^B\!H_{int}U^\dagger$, this same gathering of terms exists in the current $XY\!Z$ case, with additional bath-mediated coupling terms, as will be discussed in the following Section. The $(S_+' + S_-')$ operator with coefficient $\Delta := J_\perp$ is the standard tunnelling interaction -- instead of the usual impurity spin algebra ${\mathbf S}\rightarrow \textstyle{\frac 12}\mbox{\boldmath{$\sigma$}}$ of Pauli matrices, dressed operators $S_\pm'$ and $S_z$ (which fulfil the identical algebra) are used. 
As an aside, it should be noted that an additive contribution, corresponding to the free kinetic energy term of the infinite set of bosonic modes corresponding to the charge combination $\varphi_C(x)=(\varphi_1(x) + \varphi_2(x))/\sqrt{2}$, has been
removed.
 
By contrast with the $S_\pm'$ terms, the $S_\pm''$ terms in $U H' U^{\dagger}$ acquire exponential couplings with \emph{reinforced} strength. These unremoveable exponential terms, then, are the bosonic forms of the additional interactions brought into play by the fully anisotropic, or elliptic, form of the spin-$\frac 12$ Kondo model. 
These aspects of the model as new couplings in a generalisation of the spin-$\frac 12$ spin-boson model or two-level QDS,
will be discussed in the concluding remarks below, along with its potential interpretation as a type of quantum random walk system. We preface the discussion however with further explanation of the structure of the model and its state space.

\section{Concluding remarks: structure of the extended QDS}
\label{sec:structure}
The final step necessary in identifying the mapping to quantum dissiptive systems is the state space projection onto an eigenspace of definite total charge (total fermion number $N=N_1\!+\!N_2$). This is always possible, as the Kondo Hamiltonians are charge-preserving, ${[}H^{X\!X\!Z},N{]}=0={[}H^{XY\!Z},N{]}$. However, as is evident from the couplings, the structure of the models across the individual charge sub-sectors is quite different in each case. 

In the standard anisotropic case, $H^{X\!X\!Z}$ has a residual axial symmetry associated with conservation of the total magnetic quantum number $M = \textstyle{\frac 12}(N_1-N_2)+S_z$. This follows from the structure of the Klein operators, and their dressing of $S_\pm$ to give the effective spin operators $S_\pm'$; for example, 
${[}\textstyle{\frac 12}(N_1-N_2), {\mathcal F}_2{}^\dagger {\mathcal F}_1 {]} = - {\mathcal F}_2{}^\dagger {\mathcal F}_1$. 
Thus the system is a sum of independent models on each subspace labelled by eigenvalues of $N$ and $M$.
This means that each eigenstate of $S_z$, the spin magnetic quantum number $m_s=\pm \textstyle{\frac 12}$, occurs in a fixed charge sub-sector $(N_1,N_2)$, namely 
$(\textstyle{\frac 12}N + M - \textstyle{\frac 12},\textstyle{\frac 12}N - M + \textstyle{\frac 12})$, and
$(\textstyle{\frac 12}N + M + \textstyle{\frac 12},\textstyle{\frac 12}N - M - \textstyle{\frac 12})$, respectively. As emphasized already, each sub-sector is isomorphic to a copy of the infinite oscillator space, the dressed spin operators $S_\pm'$ together with $S_z$ map between these $m_s=\pm \textstyle{\frac 12}$ states, and we recover the standard spin-$\textstyle{\frac 12}$ spin-boson model (QDS).

\begin{figure}[!h]
\centering
\includegraphics[width=18cm,clip]{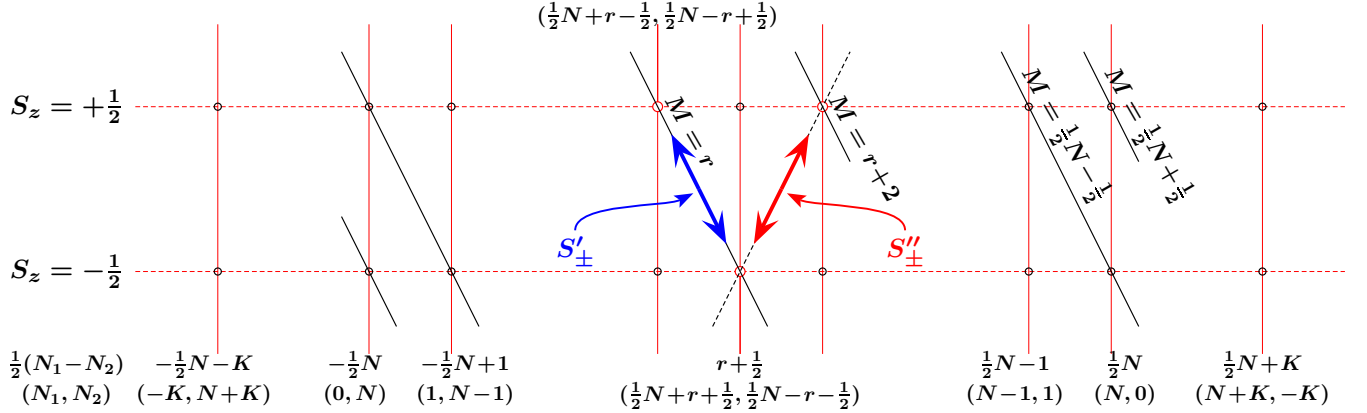}
 \caption{Diagram demonstrates the action of the dressed operators $S_\pm'$ and $S_\pm''$ on the charge sectors $(N_1,N_2)$ and spin magnetic quantum number $m_s$, for fixed total fermion number $N=(N_1+N_2)$. Standard spin raising and lowering operators give vertical shifts, while combinations of Klein operators produce horizontal shifts in $M=\textstyle{\frac 12}(N_1-N_2)+ S_z$. The diagonal lines illustrate eigenspaces of $M$, while the parameter $K$ can be arbitrarily large. The label $r$ represents an arbitrary $M$ value along the lattice and is integer for one of the two noninteracting copies of the infinitely degenerate upper and lower levels, and half-integer for the other.}
 \label{Lattice}
\end{figure}

In the fully anisotropic case, the additional terms in $H'$ ($J'\ne 0$) break the residual axial symmetry, and the quantum number $M$ is not conserved. The interpretation of the system as a type of quantum dissipative system therefore is that \emph{all} of the charge sub-sectors $(N_1,N_2)$ correlated with each spin eigenvalue constitute an extended degeneracy of these upper and lower states. Given $N_1+N_2=N$, for fixed $N$ this degeneracy is labelled by a single integer, say $N_1$. The situation is illustrated in figure \ref{Lattice} with a weight lattice enumerating the charge sectors and the accompanying spin labels. It is clear from the indicated action of the dressed operators $S_\pm'$ and $S_\pm''$ that, while $S_\pm'$ preserve with $M$, the $S_\pm''$ do not, in fact by effecting shifts of $\pm2$ units, respectively. 

The structure of the bosonised model is thus as follows. The system is an extended two-level quantum dissipative system with two noninteracting copies of infinitely degenerate upper and lower states (on even and odd states respectively), labelled by an integer `horizontal' quantum number. In addition to the standard tunnelling matrix elements between spins, and the standard dissipative couplings to $S_z$, there are additional, `bath-mediated' 
tunnelling transitions with strength $J'$ and exponential form, 
$S''_+ \exp(-i\sum_{p>0} D_p(b_{pS}+b^\dagger_{pS}))+h.c.$, for couplings $D_p = 2\sqrt{2\pi/pL}e^{-ap/2}$, where again the spin-type combinations are $b_{pS}=(b_{p1}-b_{p2})/\sqrt{2}$.


As a final comment on the potential applicability of the new QDS model as an interesting exactly-solvable system, it should be noted that the state space and interactions can be regarded as that of a type of (continuous time) quantum random walk system on the line \cite{Aharonov1993, Kempe2003}.  The `walker' part of the Hamiltonian is indeed the $H'$ piece, more recognisably written in terms of the decomposition of
$S_\pm'' = S_\pm' {\mathbb T}_\pm$ into a product of effective spin operators $S_\pm'$, and translation generators along the line, 
\[
UH'U^{-1} = J'\big(e^{-i \sum_p D_p(b_p + b^\dagger_p)} S_+' {\mathbb T}_+ +
e^{+i \sum_p D_p(b_p + b^\dagger_p)} S_-' {\mathbb T}_- \big),
\]
where ${\mathbb T}_+ = ({\mathcal F}_1{}^\dagger){}^2  ({\mathcal F}_2)^2$, ${\mathbb T}_- = ({\mathcal F}_2{}^\dagger){}^2  ({\mathcal F}_1)^2$ commute with $S_z$ but satisfy $\left[M,{\mathbb T}_{\pm}\right]=2{\mathbb T}_{\pm}$. This type of interaction characterises the `walker' evolving both in `coin' space, and undergoing horizontal displacements on the line (represented by the lattice of eigenstates of $M$; see figure \ref{Lattice}). In this case of course, as noted already, the walker steps are mediated by the additional interactions with the oscillator bath system. The further investigation of the properties of the system in the light of these connections to quantum walks, for example exploring the influence of the oscillator bath on the walker's performance, is a subject for future research. Furthermore, evaluation of quantitative aspects of this new instance of an exactly-solvable quantum dissipative system would require examination of the appropriate elliptic Bethe \emph{Ansatz} equations and their solution in terms of the physics of the dissipative system, as has been done for the trigonometric case in order to study both dynamical and thermodynamical aspects, and critical behaviour, of the spin-boson model \cite{TsvelickandWiegmann1983,Filyovetal1981,CostiandZarand1999,Takahashi1999}.

\subsection*{Acknowledgements}
We thank Prof. Ross McKenzie for discussions in the early stages of this work. This project was in part funded by the Commonwealth of Australia Endeavour Awards. 


\bibliographystyle{unsrt}

\end{document}